\documentclass{article}  
\usepackage{lajolla2006}
\usepackage{graphicx}
\frompage{000} \topage{000}                                              %|
\def\rightmark{Mini Black Holes at the LHC}
\def\leftmark{B. Betz et al.}
\title{Mini Black Holes at the LHC \\
{\small Discovery Through Di-Jet Suppression, Mono-Jet Emission and a Supersonic 
Boom in the Quark-Gluon Plasma in ALICE, ATLAS and CMS }}
\authors{
{B. Betz$^{2}$, M. Bleicher$^{2}$, U. Harbach$^{1}$, T. Humanic$^{3}$, B. Koch$^{1,2}$, H.
St\"ocker$^{1,2}$ %
}\\[2.812mm]
{\normalsize
\hspace*{-8pt}$^1$ FIAS- Frankfurt Institute for Advanced Studies\\
D--60438 Frankfurt am Main, Germany\\
\hspace*{-8pt}$^2$ Institut f\"{u}r Theoretische Physik, Johann Wolfgang Goethe -
Universit\"{a}t\\
D--60438 Frankfurt am Main, Germany\\[0.2ex] 
\hspace*{-8pt}$^3$ Department of Physics\\
Ohio State University, Columbus, OH, USA
}}
 
\abstract{We examine experimental signatures of TeV-mass black 
hole formation in heavy ion collisions at the LHC. 
We find that the black hole production results in a complete disappearance 
of all very high $p_T$ (\mbox{$> 500$} GeV) back-to-back correlated 
di-jets of total mass \mbox{$M > M_f \sim 1$}TeV. We show that the 
subsequent Hawking-decay produces multiple hard mono-jets and discuss their detection.
We study the possibility of cold black hole remnant (BHR) formation of mass $\sim M_f$
and the experimental distinguishability of scenarios with BHRs and those with complete black hole decay.
Finally we point out that a Heckler-Kapusta-Hawking plasma may form from the emitted mono-jets.
In this context we present new simulation data of Mach shocks and of
the evolution of initial conditions until the freeze-out.
}

\PACS{specifications see, e.g.\ {\tt http://www.aip.org/pacs/}}
 
\begin{document}
 
\maketitle
\setcounter{page}{1}

\section{Introduction}\label{intro}

The Frankfurt-born astronomer Karl Schwarzschild discovered the first analytic solution of the General Theory of Relativity \cite{Schwarzschild}. 
He layed the ground for studies of some of the most fascinating 
and mysterious objects in the universe: 
the black holes. Recently, it was conjectured that black holes (BH) do also reach into the regime of particle and collider physics: 
In the presence of additional
compactified large extra dimensions (LXDs), it seems possible to produce tiny 
black holes in colliders such as the Large Hadron 
Collider (LHC), at the European Center for Nuclear Research, CERN. 
This would allow for tests of Planck-scale physics and of the onset 
of quantum gravity - in the laboratory! Understanding black hole physics is a key to the phenomenology of these new effects beyond the 
Standard Model (SM).

During the last decade, several models \cite{Antoniadis:1990ew,add,rs1} using extra dimensions as an additional assumption to the 
quantum
field theories of the Standard Model (SM) have been proposed. The most intriguing feature of these models is that they provide a 
solution
to the so-called hierarchy problem by identifying the "observed" huge 
Planck-scale as a geometrical feature of the space-time,
while the true fundamental scale of gravity $M_f$ may be as low as 1 TeV. 
The setup of these effective models is partly motivated by String
Theory. The question whether our space-time has additional dimensions is well-founded on its own and worth the effort of 
examination. 

In our further discussion, we use the model proposed by Arkani-Hamed, Dimopoulos and Dvali
\cite{add}, proposing $d$ extra spacelike dimensions without curvature, each of them compactified to a certain radius $R$. 
Here all SM 
particles are confined to our 3+1-dimensional brane, while gravitons are allowed to propagate freely in the (3+d)+1-dimensional bulk. 
The Planck mass $m_{Pl}$ and the fundamental mass $M_f$ are related by
\begin{eqnarray}
m_{Pl}^2 = M_f^{d+2} R^d \quad. \label{Master}
\end{eqnarray}

The radius $R$ of these extra dimensions can be estimated using Eq.(\ref{Master}). 
For $d$ eqaling $2$ to $7$ and $M_f \sim$~TeV, $R$ extends from $2$~mm to $\sim 10$~fm.
Therefore, the inverse compactification radius $1/R$ lies in energy range 
eV to MeV, respectively. The case $d=1$ is excluded: It would result in an extra dimension about the size of the solar system. For 
recent 
updates on constraints on the parameters $d$ and $M_f$ see e.g. Ref.\ \cite{Cheung:2004ab}.

\section{Estimates of LXD-black hole formation cross sections at the LHC}

The most exciting signature of {\sc LXD}s is the possibility of black hole production in colliders 
\cite{Banks:1999gd,dim,ehm,Giddings3,Hofmann:2001pz,Hossenfelder:2001dn,Hossenfelder:2003jz,Hossenfelder:2003dy,Casadio:2001dc,Alexeyev:2002tg,BleicherNeu,Chamblin:2002ad,Casanova:2005id,own2,Casadio:2001wh,Hossenfelder:2004ze,Hossenfelder:2005bd,Hst06,Hst062,Alberghi:2006qr} 
and 
in ultra high energetic cosmic ray events \cite{cosmicrayskk,cosmicraysbh}: At distances below the size of the extra
dimensions the Schwarzschild radius \cite{my} is given by
\begin{equation} \label{ssradD}
R_H^{d+1}=
\frac{2}{d+1}\left(\frac{1}{M_{f}}\right)^{d+1} \; \frac{M}{M_{f}}
\quad .
\end{equation}
This radius is much larger than the corresponding radius in 3+1 dimensions. Accordingly,
the impact parameter at which colliding particles form a black hole via the 
Hoop conjecture \cite{hoop} rises enormously in the extra-dimensional
setup.
The LXD-black hole production cross section can be approximated by the classical geometric 
cross section
\begin{eqnarray} \label{cross}
\sigma(M)\approx \pi R_H^2 \quad,
\end{eqnarray}
which only contains the fundamental scale as a coupling constant.

This classical cross section has been under debate
\cite{Voloshin:2001fe,Rychkov:2004sf,Jevicki:2002fq}: Semi-classical considerations 
yield form factors of order one 
\cite{Formfactors}, which take into account that only a fraction of the
initial energy can be captured behind the Schwarzschild-horizon. 
Angular momentum 
%$J  \approx 1/2 M b$ 
considerations change the results by a factor 
of two \cite{Solo}. Nevertheless, the naive classical result 
remains valid also in String Theory \cite{Polchi}.

\begin{figure}[htb]
\begin{minipage}[c]{6 cm}
\vspace*{-.1cm}
\begin{center}
\includegraphics[width=5.5truecm]{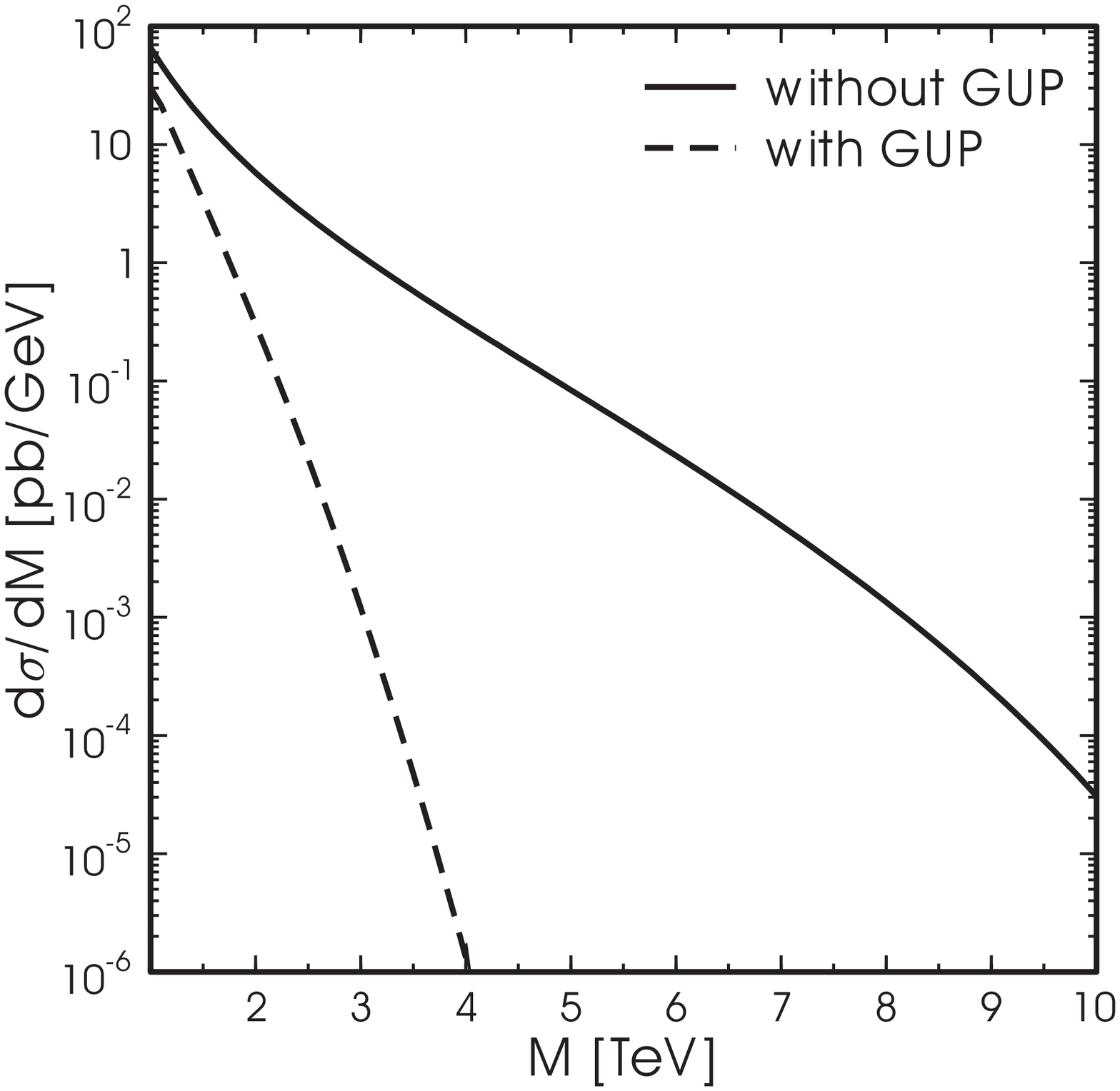}
\end{center}
\end{minipage}
\begin{minipage}[c]{6 cm}
\vspace*{-.1cm}
\begin{center}
\includegraphics[width=5.5truecm]{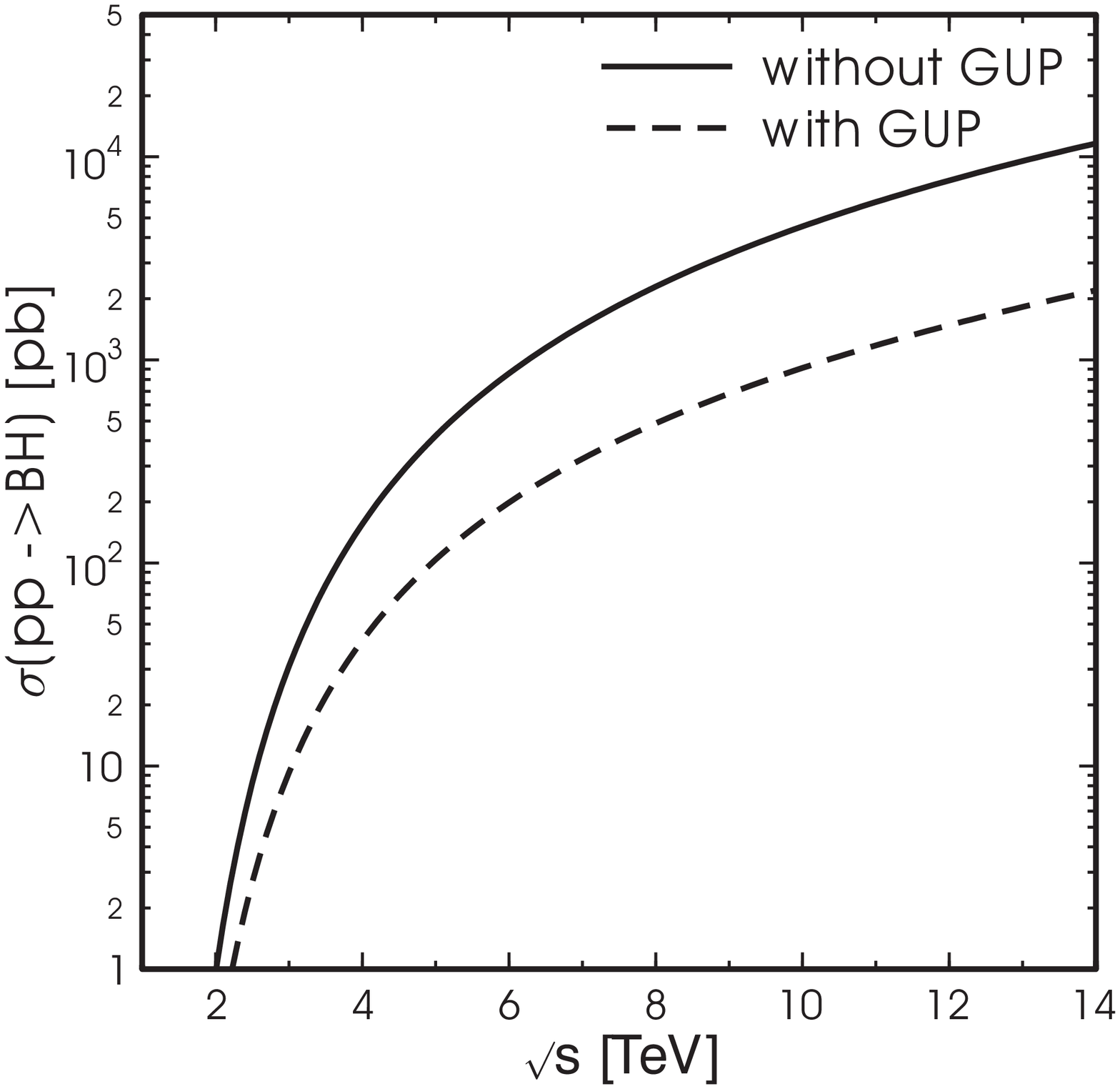}
\end{center}
\end{minipage}
\caption{The left plot shows the differential cross section for black hole
production in p-p collisions at $\sqrt{s}=14\rm{TeV}$ ({\sc LHC}) for $M_{f}=1$~TeV. 
The right plot shows the integrated cross section for 
BH production as a function of the 
collision energy $\sqrt{s}$. In both cases, the curves for various
$d$ differ only slightly from the above depicted ones.
The dashed curves show calculations including the minimal length
(via a Generalized Uncertainty Principle (GUP)) {\protect\cite{Hossenfelder:2004ze,Hossenfelder:2005bd}}.}
\protect{\label{dsdm}}
\end{figure}

Stronger modifications to the BH cross section are expected from recent calculations introducing a minimal length scale,
suggested by String Theory and Loop Quantum Gravity alike. Via the use of a model implementing a
Generalized Uncertainty Principle (GUP), one can show 
that a minimal length scale leads to a reduction of the density 
of states in momentum space at high energies. The squeezing of the
momentum states not only reduces the 
black hole cross section, but also Standard Model cross sections involving high momentum transfer
\cite{Hossenfelder:2004ze}, see Fig. \ref{dsdm}.
\vspace*{1ex}
%#############################################################################

Setting $M_f\sim 1~$TeV and $d=2 - 7$ one finds cross-sections of $\sigma \sim 400$~pb$ - 10$~nb. Using the geometrical 
cross section formula, it is now possible to compute the differential cross section ${\mathrm d}\sigma/{\mathrm d}M$ for p-p collisions 
with an invariant energy $\sqrt{s}$. This cross section is given by the
summation over all possible parton interactions and integration over the momentum fractions $x_i$, where  the kinematic relation 
$x_1 x_2 s=\hat{s}=M^2$ has to be fulfilled. This yields the expression
\begin{eqnarray} \label{partcross}
\frac{{\rm d}\sigma}{{\rm d}M}
&=&  \sum_{A, B} \int_{0}^{1} {\rm d} x_1 \frac{2 \sqrt{\hat{s}}}{x_1s}
f_A(x_1,\hat{s})
f_B (x_2,\hat{s})  \sigma(M,d).   \quad
\end{eqnarray}

A numerical evaluation \cite{Hossenfelder:2005bd} using the 
{\sc CTEQ4}-parton distributions $f_i(x,Q)$ results in the cross section 
displayed in Figure \ref{dsdm}. 

One can see that independent of the specific scenario, most of the black holes created have masses 
close to the production threshold. This is due to the fact that the parton distribution functions $f_i(x_i)$ are strongly peaked at 
small values of the momentum fractions $x_i$.
%\begin{figure*}[htb]
%\vspace*{-.8cm}
%                \insertplot{sigmapp.eps}
%\vspace*{-2cm}
%\caption{The left plot shows the differential cross section for black hole
%production in proton-proton-collisions at the {\sc LHC} for $M_{f}=1$~TeV. 
%The right plot shows the integrated total cross section as a function of the 
%collision energy $\sqrt{s}$. In both cases, the curves for various
%$d$ differ from the above depicted ones by less than a factor
%10 and the lower curves show the calculations with minimal length included.\protect\cite{Hossenfelder:2004ze,Hossenfelder:2005bd}}
%\protect{\label{dsdm}}
%\end{figure*}

At the LHC up to $10^9$ black holes may be created per year with the 
estimated full LHC luminosity of $L=10^{34}{\rm cm}^{-2}{\rm s}^{-1}$ at $\sqrt{s}=14$~TeV: Depending on the specific scenario, about 
ten black holes per second could be created \cite{dim}.
LXD-black hole production would have dramatic consequences for future collider physics: Once the collision energy crosses the threshold 
for black hole production, no further information about the structure of matter at small scales can be extracted - 
this would be ''{\it{the end of short distance physics}}'' \cite{Giddings3}.

%#############################################################################

\section {Suppression of high mass correlated di-jet signals in heavy ion collisions}

The above findings led to a high number of publications on the topic of 
TeV-mass black holes at colliders
\cite{dim,ehm,Giddings3,Hofmann:2001pz,Hossenfelder:2001dn,Chamblin:2002ad,own2,Casadio:2001wh,Hst06,Hst062,Landsberg:2002sa,Marcus,Lonnblad:2005ah,Hossenfelder:2005ku,Humanic,PYTHIA,CHARYBDIS,BHex,Koch:2005ks}, 
for hadronic collisions as well as for heavy ion 
collisions \cite{own2,Hst06,Chamblin:2003wg}: 
At the same center of mass energy, the number of black holes in a heavy ion event 
compared to a hadronic event is increased by about
thousandfold due to the scaling with the number of binary collisions \cite{Chamblin:2003wg}.
We will discuss special features of heavy ion collisions in Sec.\ \ref{hi}.

The first, cleanest signal for LXD-black hole formation in Pb-Pb collisions 
is the complete suppression of high energy back-to-back-correlated 
di-jets with $M > M_f$: two very high energy partons
which usually define the di-jets in the Standard Model, each having an energy of 
$\sim$ one-half $M_f$ (i.e. $p_T \geq 500$ GeV), now end up inside the black 
hole \cite{own2,Casadio:2001wh,Hst06,Lonnblad:2005ah} instead of being observable 
in the detector.
Di-jets with $E_{di-jet} > M_f$ cannot be emitted.

%%%%%%%%%%%%%%%%%%%%%%%%%%%%%%%%%%%%%%%%%%%%%%%%%%%%%%%%%%%%%%%

\section{Hard, isotropic multiple mono-jet emission as a signal for hot LXD-black hole hawking-evaporation}

Once produced, the black holes may undergo an evaporation process
\cite{Hawk1}
whose thermal properties carry information about the parameters $M_{\rm f}$ and $d$. An
analysis of the evaporation
will therefore offer the possibility to extract knowledge about the
topology of space time and the underlying theory.

The evaporation process can be categorized in three characteristic stages
\cite{Giddings3,Hossenfelder:2001dn,Hossenfelder:2005ku}:

\begin{enumerate}
\item {\sc Balding phase:} In this phase the black hole radiates away the
multipole
moments it has inherited
from the initial configuration and settles down in
a hairless state. During this stage, a certain fraction 
of the initial mass will be
lost in gravitational
radiation.
\item {\sc Evaporation phase:} The evaporation phase 
starts with a spin
down phase
in which the Hawking
radiation carries away the angular momentum, after which it proceeds with
emission
of thermally
distributed quanta until the black hole reaches Planck mass. The radiation
spectrum
contains
all Standard Model (and possibly SUSY-) particles, which are emitted on our brane, as well as
gravitons,
which may also propagate into the extra dimensions as well. It is expected that most of the initial
energy is emitted during this phase into Standard Model particles.
A very thorough description of these evaporation characteristics has been
given in \cite{Kanti:2004nr}.
\item {\sc Planck phase:} Once the black hole has reached a mass close to
the Planck
mass, it falls into
the regime of quantum gravity and predictions become increasingly
difficult. It is
generally
assumed that the black hole will then either completely decay in a few
Standard
Model particles or
form a quasi-stable remnant.
\end{enumerate}

To understand the signature caused by black hole decay,
we have to examine the Hawking evaporation process in detail:
The evaporation rate ${\rm d}M/{\rm d}t$ can be
computed for an arbitrary number of dimensions using the thermodynamics of black holes. 
The Hawking-temperature ($T$) depends on the black hole radius
\begin{eqnarray} \label{tempD}
T=\frac{1+d}{4 \pi}\frac{1}{R_H} \quad,
\end{eqnarray}
which is given by Eq. (\ref{ssradD}).
The smaller the black hole, the larger its temperature.

\begin{figure*}[htb]
\vspace*{-.8cm}
                 \insertplot{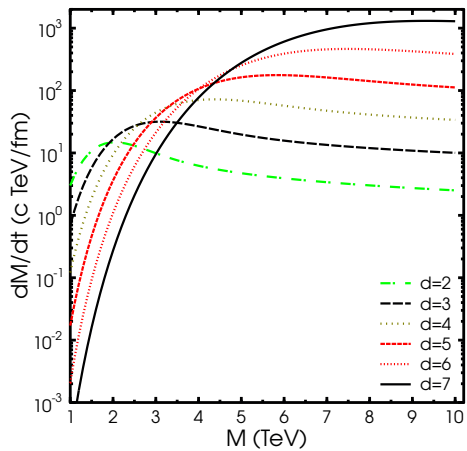}
\vspace*{-1.2cm}
\caption{The Black hole evaporation rates as a function of the initial black hole mass for various $d$ \cite{Hossenfelder:2001dn}. 
\label{mdot}}
\end{figure*}

Integrating the thermodynamic identity d$S/$d$M = 1/T$ over $M$ yields 
the entropy
\begin{eqnarray}
S(M)
&=& 2 \pi  \frac{d+1}{d+2} \left( M_f R_H \right)^{d+2}\quad.
\end{eqnarray}
With rising temperature,
the emission of a particle will have a non-negligible
influence on
the total energy of the black hole. This problem can appropriately be
addressed by
including the back-reaction of the emitted quanta as derived in Ref.\ \cite{Page,backreaction}. It is found that in the regime of
interest, when
$M$ is of order $M_{f}$, the number density
for a single particle micro state $n(\omega)$ is modified and now given by the change
of the black
hole's entropy:
\begin{equation} \label{nsingle}
n(\omega) =  \frac{\exp[S(M-\omega)]}{\exp[S(M)]}\quad .
\end{equation}
From this, using the evaporation rate we obtain
\begin{eqnarray} \label{mdoteq}
\frac{{\mathrm d}M}{{\mathrm d}t} = \frac{\Omega_{(3)}^2}{(2\pi)^{3}}
R_H^{2} 
\int_{0}^{M} 
\frac{\omega^3 \,\rm{d}\omega}{\rm{exp}\left[S(M-\omega)-S(M)\right]}\quad,
\end{eqnarray}
where $\Omega_{(3)}$ is the 3-dimensional unit sphere.
Fig. \ref{mdot}
shows this rate as a function of M 
for various $d$. 
One observes that the evaporation process
of the black holes slows down
in its late stages \cite{Hossenfelder:2001dn,Hossenfelder:2003jz}
\footnote{In a 3-dimensional theory this enhanced lifetime can
also be obtained from a renormalization group 
approach \cite{Bonanno:2006eu}.}, 
and may even come to a complete stop, thus, stable black hole remnants may be 
formed 
\cite{Hossenfelder:2003jz,Hst06,Hst062,Bonanno:2006eu,Koch:2005ks}.

%
%#############################################################################

%#############################################################################

The above discussion allows for the following observations:
\begin{itemize}
\item Typical temperatures for LXD-black holes with $M_{BH} \gg M_f$, e.g. $5-10$~TeV,
are several hundred GeV. This high temperature results in a very short lifetime. The black
hole will decay close to the primary interaction region and thus its decay products can be observed
in collider detectors.
\item Most of the SM particles of the black body radiation are emitted with $\sim 100$~GeV average
energy, which leads to multiple high energy mono-jets with much higher multiplicity than in Standard Model
processes \cite{Hst06}.
\item The total number of emitted jets can be estimated to be of order $10$. Because of the thermal
characteristics of the decay, the pattern will be nearly isotropic, with  a high sphericity of the event.
\end{itemize}

Ideally, the energy distribution of the decay products allows for a
determination of the temperature (by fitting the energy spectrum to the
predicted shape) as well as of the total mass of the BH (by summing up all
energies).
This then will allow for a reconstruction of the fundamental scale $M_{\rm f}$ and the
number of extra dimensions.

Several experimental
groups have included LXD-black hole searches into their research programs for
physics beyond the Standard Model, in particular the ALICE, ATLAS and 
CMS collaborations at the LHC \cite{Humanic}. PYTHIA 6.2 \cite{PYTHIA} 
with the CHARYBDIS \cite{CHARYBDIS} event generator allows for a simulation of
black hole events and data reconstruction from the
decay products. Such analysis has been summarized in Refs. \cite{Humanic,Atlas,BHex}.

\section{Formation of stable black hole remnants and single track detection in the ALICE-TPC}

%\begin{figure}[htb!]
%\vspace*{-.8cm}
%                 \insertplot{mtrel2.eps}
%\vspace*{-1.3cm}
%\caption{The mass evolution for a black hole of initial mass $M=10$~TeV and
%various $d$. Here, we set $M_{\rm R} = M_{\rm f} =1$~TeV.}
%\label{mtrel2}
%\end{figure}
To obtain predictions for collider experiments, one has to produce numerical simulations incorporating black hole events. 
These simulations have been performed but have so far assumed mostly that the black holes decay completely into Standard 
Model particles. 
As already pointed out, however, there are equally strong indications that the black holes do NOT
evaporate completely, but rather leave a meta-stable black hole remnant
(BHR) \cite{Hossenfelder:2001dn,Hossenfelder:2003jz,Hossenfelder:2003dy,Hst06,Hst062,Bonanno:2006eu,Koch:2005ks}.

%
%Fig.~\ref{mtrel2} shows that this is not the case: the mass evolution of
%the  produced black holes stabilizes rapidly, $t < 1$fm/c, the average
%energy of emitted particles drops to zero within $10$ fm.
%
\begin{figure}[htb]
\vspace*{-.8cm}
                 \insertplot{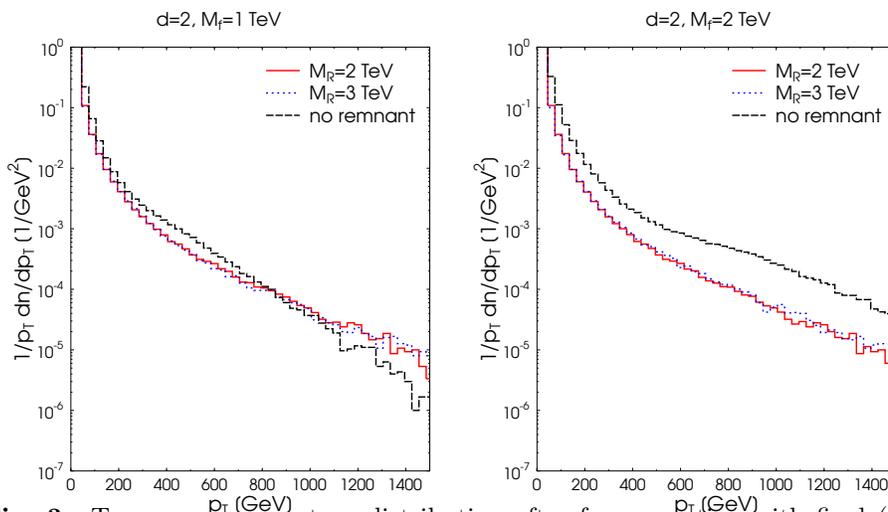}
\vspace*{-1.cm}
\caption{Transverse momentum distribution after fragmentation with
final (two-body) decay in contrast to the formation of a black hole
remnant \protect\cite{Koch:2005ks}}.
\label{ptd2}
\end{figure}

Figure \ref{ptd2} shows the $p_T$-spectra after fragmentation for complete black hole decay as well
as for remnant formation for p-p collisions at the LHC. The spectra have been obtained with the CHARYBDIS simulation 
\cite{CHARYBDIS} and the observables computed within the PYTHIA framework.
It is apparent that the additional contribution from the final decay causes
a clear difference between the curves with/without remnant formation.
The BHR signal is thus clearly distinguishable from
disappearing black holes.\footnote{The graph does not include background, but 
 detectors like ALICE can differentiate 
LXD-black holes from the QCD background \cite{Humanic,Cortese:2002kf}.}
These results also agree very well with analytically computed results
\cite{Koch:2005ks}.

\begin{figure*}[htb]
%\vspace*{-.4cm}
                 \insertplot{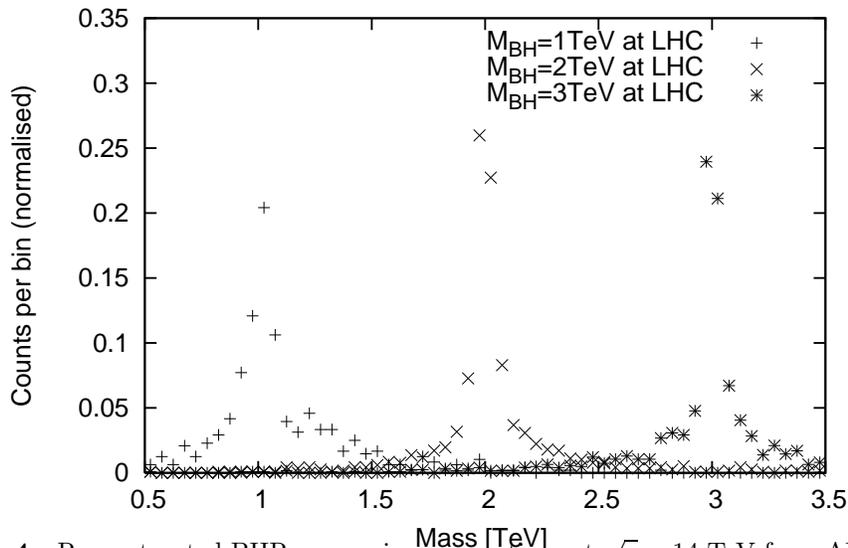}
\vspace*{-1.cm}
\caption{
Reconstructed BHR masses in p-p reactions at $\sqrt{s}=14$ TeV from ALICE (TOF $56$ ps) resolution
for $M_{\rm BH}=1,2,3$ TeV.\label{Bhm}
}
\end{figure*}

If BHRs are formed, they can carry charge and may thus not only be reconstructed via decay products, but can rather 
directly be observed: 
Charged BHRs should appear in 
the ALICE detector at the LHC as a magnetically very
stiff charged (small curvature) track.
As shown in Fig. \ref{Bhm}, the mass of a charged BHR can be reconstructed
within the ALICE time of flight and spatial resolution \cite{Humanic}. 

\section{Heckler-Kapusta-Hawking plasma and hard mono-jet suppression}

The energy density of the multiple Hawking mono-jets emitted from the evaporating black holes is enormous: Several TeV are emitted 
within a 4-sphere of $\sim (10^{-1})^4 - (10^{-2})^4\, \hbox{fm}^4$/c, implying energy densities of
$10^6$ to $10^9\, \hbox{GeV fm}^{-3}$, i.e. many orders of magnitude higher than the energy densities expected for the 
quark-gluon plasma ($\epsilon\sim 500 \,\hbox{GeV fm}^{-3}$) which is expected to be created in Pb-Pb collisions at the LHC at 
$\sqrt{s}_{NN}=5.5$~TeV \cite{Hst06}.

Hence, the question arises whether the multiple jets thermalize at this enormous energy density and form an ultra-hot 
($T\gg T_{EW}\gg T_{QCD}$) plasma of Standard Model- (plus SUSY-)particles.

Such a hot Heckler-Kapusta-Hawking (HKH) plasma-scenario for primordial 3+1-dimensional black 
holes which could now decay and show up in cosmic radiation components has been studied by Heckler, Kapusta and co-workers 
\cite{Heckler:1995qq,Kapusta:2000xt}. 
For the LXD-black holes and BHRs to be created in p-p reactions at the LHC, 
a similar plasma might be produced \cite{Hofmann:2001pz}: Such a HKH plasma at $T\sim 1$~TeV should contain many massless SM particles, 
as $T$ is above the electroweak phase transition temperature \cite{Hst06,Hst062}. 
The bare masses of e.g. $W^{+/-}$ and $Z$ as well as light supersymmetric partners 
may thus become accessible to experiment. Further interesting topics to
study in this context include thermalization and viscosity of this SM-SUSY state of matter, hydrodynamic expansion, abundant 
pre-hadronic freeze-out emission of
(otherwise rare) SUSY particles, quarks $(b,t)$ and leptons \cite{Hst06,Hst062}.

After formation, the HKH plasma will expand rapidly according to the relativistic viscous hydrodynamic
transport equations \cite{Csernai:1980sc}.
In the ideal fluid approximation, this three-dimensional radial expansion can be
approximated by the simple relativistic blast wave model \cite{Siemens:1978pb,Stoecker:1981za},
which decribes the particle spectra
\begin{equation}
\frac{{\rm d}^3\sigma}{{\rm d}p^3}=N\exp(-\gamma E/T)
\bigg[\bigg(\gamma+\frac{T}{E}\bigg)\frac{\sinh{\alpha}}{\alpha}-\frac{T}{E}\cosh{\alpha}\bigg]\quad ,
\end{equation}
after a common hydrodynamic
isentropic expansion freezing out with a common freeze-out temperature $T_{f.o.}$,
and a common collective freeze-out velocity $v_{f.o.}$.\footnote{Here, $\gamma=(1-v_{f.o.}^2)^{-1/2}$, 
$\alpha=\gamma v_{f.o.} p/T$ and $N$ gives an absolute normalization.}

The nearly isentropic radial expansion causes strong space-momentum correlations, resulting in a spherical 
shell of "cool" (i.e. $T=150$ MeV) hadronic matter at the hadronic freeze-out. Nearly all initial 
thermal energy is then transformed into collective radial motion, i.e. the mean flow velocity $v_F$ is close to the speed of
light. Hence, the invariant $p_T$ spectra of hadrons exhibit peaks at the $p_T=100$ GeV range. The residual hadronization temperature of 
$T=150$~MeV in the rest
frame of the fluid causes small thermal smearing in the spectra.
We see that the final spectra seen in the HKH plasma scenario will appear quite differently than the shoulder-arm spectra in heavy ion 
collisions, which exhibit moderate flow velocities of
$v_F = 0.5\,c$. 

\section{Black holes created in heavy ion collisions: What are the interactions of BHR, the hot HKH plasma and of Hawking-radiation
jets with the quark-gluon plasma created in Pb-Pb at the LHC?}\label{hi}

In p-p collisions, the above discussed HKH fluid expansion and/or Hawking-radiation jet-emission will take place in the vacuum.
In heavy ion collisions, black holes created in individual parton-parton collisions will create
the hot HKH fluid, which expands {\bf "into"} the surronding -- much cooler -- 
QCD plasma which has been created by soft parton collisions of up to 400
participating nucleons with an initial temperature of $T\sim 500$~MeV, i.e. about 1/1000-th
of the Black Hole plasma temperature, $T\sim 500$~GeV. 
The extremely high energy density in the HKH plasma can cause shock discontinuities, travelling as nonlinear high density shock waves 
through the quark-gluon plasma.
This phenomenon is quite analogous to the shock fronts discussed in heavy ion collisions since three decades 
\cite{shockfronts,Stoecker:2004qu,Satarov:2005mv}.

A recent interesting speculation about the fate of hard jets emitted from the hard parton-parton collisions in Pb-Pb reactions (both at
RHIC and LHC) has been the prediction of Mach-shock cones, plasma wakes or, respectively, curved Mach shock waves, excited in the 
dense medium by the propagating jets \cite{Hst06,Stoecker:2004qu,Satarov:2005mv,Betzinprog}.

The properties of the matter in front and behind such shockfronts can be calculated analytically in the
planar approximation using the Rankine-Hugoniot-Taub adiabate (RHTA)\cite{shockfronts}
\begin{eqnarray}
\label{RHTA}
\frac{w_1^2}{\rho_1^2}-\frac{w_2^2}{\rho_2^2}+
(p_2-p_1)\left(\frac{w_1}{\rho_1^2}+\frac{w_2}{\rho_2^2}\right)=0\quad,
\end{eqnarray}
where $w=e+p$ denotes the enthalpy, $\rho$ the particle density, and $p$ the pressure in the medium in front (1) and behind (2) the
relativistic shockfront.
\begin{figure*}[htb!]
%\vspace*{-.8cm}
                 \insertplot{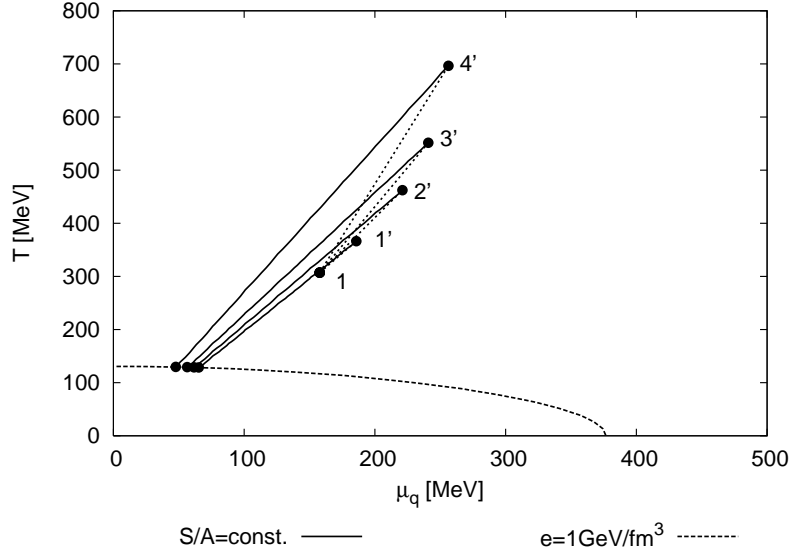}
%\vspace*{-2cm}
\caption{Jet-induced secondary hydrodynamic shocks travelling through the baryonrich QGP 
(1) with $\epsilon_1=20\,\hbox{GeV/fm}^3$, $\rho_1= 2\, \hbox{fm}^{-3}$ 
$S/A=41$. 
Four different $e_2$-values are adopted: (1') $\epsilon_2=40\, \hbox{GeV/fm}^3$, $S/A=41.6$, 
(2') $\epsilon_2=100\, \hbox{GeV/fm}^3$, $S/A=44$, 
(3') $\epsilon_2=200\, \hbox{GeV/fm}^3$, $S/A=48.1$, (4') $\epsilon_2=500\, \hbox{GeV/fm}^3$, $S/A=57$. }
\label{dshock1}
\end{figure*}

Fig. \ref{dshock1} shows shock waves calculated from Eq.\ (\ref{RHTA}) with different energy densities  
$\epsilon_2=40, 100, 200$ and  $400\,\hbox{GeV/fm}^3$ that travel through the "normal", 
heavy ion induced quark-gluon plasma which is assumed to have an energy density of $\epsilon_1=20\,\hbox{GeV/fm}^3$.
The isentropes (full lines) lead to the respective freeze-out points using a hadronization criterion of
$\epsilon=1\,\hbox{GeV/fm}^3$. Note that the jet-induced shocks create considerable entropy. This leads to a strong shift of the
hadronization density and baryochemical potential per quark, $\mu_q$, to lower values.

Now, a Heckler-Kapusta-Hawking plasma with an initial temperature of $500$~GeV can explode into the
quark-gluon plasma with a temperature of $\sim 500$~MeV. As mentioned above, this hot HKH plasma 
contains besides the different quarks and leptons also photons, electrons, myons, tauons, $W^{+/-}$ and Z bosons, whereas
the quark-gluon plasma created during the heavy ion collision is formed by quarks and gluons. Therefore, additional 
conservation laws have to be taken into account at the shock front \cite{Betzinprog}. The 
characteristics of the jet-induced hydrodynamic shocks is analogous to Fig.\ \ref{dshock1}. However, the hadronization chemical
potential is ignorable small for all practial purposes. Hence, exact particle-antiparticle symmetry should prevail, 
just as in the early universe.

Mach shocks can also occur if the HKH plasma is not created, but instead mono-jets from black holes (formed in 
Pb-Pb collisions at the LHC) induce Mach-waves travelling through the 
quark-gluon plasma, as calculated in Fig.\ \ref{RHTA}. They can easily be observed, as at RHIC, by the two- and three- particle 
correlation functions, at Mach angles given by the well-known Mach relation $\phi = \arccos(v_s/v_{jet})$, where $\phi$ is the emission 
angle of the Mach shock particles relative to the jet axis.

A final word to the propagation of a possible black hole remnant (BHR) through the quark-gluon plasma: the average momentum of such a BHR 
is huge (several hundred GeV), and despite the expected mass $M_{\rm BHR} = M_f = 1$ TeV, the typical BHR-velocity moving through the
quark-gluon plasma is 
calculated to be about $v_{\rm BHR}\sim0.9c$, at least for the lightest black holes.
Hence, BHR-induced Mach cones may be caused by the relative flow of the Pb-Pb QCD plasma relative to the BHR can occur for either
\begin{itemize}
\item fast BHRs with velocity $v_{\rm BHR} \gg c_{QGP} = 0.57c$, which are produced at the  center of the QGP, where the flow of 
the 
plasma relative to the center-of-mass frame is negligible or
\item slower BHRs which are created in the periphery of the Pb-Pb corona, but move "inwards", against the strong outward flow of 
the Pb-Pb 
QCD-plasma, for which the expected average outward "radial" flow velocities may considerably exceed $c_s \sim 0.57c$.
\end{itemize}

\section{Conclusion}
The LHC will provide exciting discovery potential way beyond supersymmetric extensions of the Standard Model. We have shown the 
different aspects of black hole formation and decay on microscopic scales and have discussed experimental signatures. In particular, we 
predict a complete suppression of back-to-back correlated di-jets, independent of the specific scenario. We have discussed the 
possibility of the formation of stable black hole remnants and have shown how signatures in the ALICE TPC chamber can be used to 
discriminate between complete decay and remnant formation. We also examine the possibility of the formation and expansion of a 
Heckler-Kapusta-Hawking plasma outside the horizon of the black hole. Without such a plasma, hard mono-jets from black hole decay would 
be observed in the 
detector; with such a plasma, those mono-jets will be absorbed and replaced 
by a high number of energtic particles expanding in a blast wave. In the case of black hole 
formation in a heavy ion collision, additional effects are observable due to the interaction of the black hole products
with the surrounding bulk quark-gluon plasma. In particular, 
Mach shocks in the QGP may be caused by either the blast wave from the HKH plasma or by
hard mono-jets from black hole decay or by fast moving black hole remnants.

\section*{Acknowledgements}
This work has been supported by GSI, BMBF and the ALICE collaboration.

Discussions with and important contributions by the following colleagues are acknowledged:

Harry Appelsh\"auser, Peter Braun-Munzinger, 
Adrian Dumitru, Sabine Hossenfelder, Kerstin Paech and Gebhard Zeeb.


\begin{thebibliography}{1}
\expandafter\ifx\csname url\endcsname\relax
  \def\url#1{{\tt #1}}\fi
\expandafter\ifx\csname urlprefix\endcsname\relax\def\urlprefix{URL }\fi

\bibitem{bib1}
A.~Coalman, D.~Fiedler and G.~Hughes, {\it Rev. Mod. Phys.\/} {\bf 63} (1991)
  545.

\bibitem{bib2}
G.~Banderer and S.~Mikhalchuk, {\it Phys. Lett.\/} {\bf B54} (1983) 387; {\it
  Sov. J. Nucl. Phys.} {\bf 19} (1979) 197.

\bibitem{bib3}
D.~Heinemann and U.~S{\o}renmann, {\it Phys. Rev. Lett.\/}  (2001).
  [hep-ph/0123456].

\bibitem{bib4}
T.~Czukor and J.~Zel\'enyi, {\it Acta Phys. Hung. A\/} {\bf 24} (2005) 345.

\bibitem{bib5}
I.~Mileev, K.~Plummer and M.~Gorentchev, {\it Int. J. Mod. Phys.\/} {\bf A11}
  (1995) 77.

\bibitem{bib6}
S. Trottman and G. Karib, {\it Proceedings of the 5th Workshop on High Energy
  Collision Phenomena}, August 2--6, 1993, eds. J.-P. Condrieu and
  P.~Ch\^{a}teau-Simone, 1994, Springer, p. 324.

\bibitem{bib7}
W.T. Fogg, Talk presented at {\it Advanced Research Workshop on Hot Hadronic
  Matter: Theory and Experiment}, 1994, Barcelona, June 29 -- July 1, to be
  published in the Proceedings by Plenum Press.

\bibitem{bib8}
This was not seen in [6].

\bibitem{bib9}
From Eq.~(\ref{eq2}) one sees that in the original frame the cross term is
  given by: $$ Q_{\perp L}^2 = \left( <x\;\tau\;{\rm sh}\eta'>- <x><\tau\;{\rm
  sh}\eta'>\right) $$ up to small logarithmic corrections. This clearly
  vanishes as soon as the source becomes reflection symmetric under $\eta' = -
  \eta'$.

\end{thebibliography}


\begin{thebibliography}{0}

\bibitem{Schwarzschild}
K.~Schwarzschild, Sitzungsberichte der Deutschen Akademie der Wissenschaften zu Berlin, Klasse f\"ur
Mathematik, Physik und Technik, 189 (1916)

\bibitem{Antoniadis:1990ew} 
I.~Antoniadis, 
%``A Possible New Dimension At A Few Tev,'' 
Phys.\ Lett.\ B {\bf 246}, 377 (1990); 
%%CITATION = PHLTA,B246,377;%% 
%\cite{Antoniadis:1996hk} 
%\bibitem{Antoniadis:1996hk} 
I.~Antoniadis and M.~Quiros, 
%``Large radii and string unification,'' 
Phys.\ Lett.\ B {\bf 392}, 61 (1997) 
[arXiv:hep-th/9609209];
%\bibitem{Dienes:1998vg} 
K.~R.~Dienes, E.~Dudas and T.~Gherghetta, 
%``Grand unification at intermediate mass scales through extra dimensions,'' 
Nucl.\ Phys.\ B {\bf 537}, 47 (1999) 
[arXiv:hep-ph/9806292];
%%CITATION = HEP-PH 9806292;%% 
%\cite{Dienes:1998vh} 
%\bibitem{Dienes:1998vh} 
K.~R.~Dienes, E.~Dudas and T.~Gherghetta, 
%``Extra spacetime dimensions and unification,'' 
Phys.\ Lett.\ B {\bf 436}, 55 (1998)
 [arXiv:hep-ph/9803466].
%%CITATION = HEP-PH 9803466;%% 
   
\bibitem{add} 
N.~Arkani-Hamed, S.~Dimopoulos and G.~R.~Dvali,
%``The hierarchy problem and new dimensions at a millimeter,''
Phys.\ Lett.\ B {\bf 429}, 263 (1998)
[arXiv:hep-ph/9803315];
%%CITATION = HEP-PH 9803315;%%
 I.~Antoniadis, N.~Arkani-Hamed, S.~Dimopoulos and G.~R.~Dvali,
%``New dimensions at a millimeter to a Fermi and superstrings at a TeV,''
Phys.\ Lett.\ B {\bf 436}, 257 (1998)
[arXiv:hep-ph/9804398];
%%CITATION = HEP-PH 9804398;%% 
N.~Arkani-Hamed, S.~Dimopoulos and G.~R.~Dvali,
%``Phenomenology, astrophysics and cosmology of theories with  sub-millimeter
%dimensions and TeV scale quantum gravity,''
Phys.\ Rev.\ D {\bf 59}, 086004 (1999)
[arXiv:hep-ph/9807344].
%%CITATION = HEP-PH 9807344;%%


\bibitem{rs1}  
%\cite{Randall:1999vf}
%\bibitem{Randall:1999vf}
L.~Randall and R.~Sundrum,
%``An alternative to compactification,''
Phys.\ Rev.\ Lett.\  {\bf 83}, 4690 (1999)
[arXiv:hep-th/9906064];
%%CITATION = HEP-TH 9906064;%%L.~Randall \& R.~Sundrum, 
%and 
%\cite{Randall:1999ee}
%\bibitem{Randall:1999ee}
%L.~Randall and R.~Sundrum,
%``A large mass hierarchy from a small extra dimension,''
Phys.\ Rev.\ Lett.\  {\bf 83}, 3370 (1999)
[arXiv:hep-ph/9905221].
%%CITATION = HEP-PH 9905221;%%L.~Randall \& R.~Sundrum, 
 

\bibitem{Cheung:2004ab}
K.~Cheung
%``Collider phenomenology for a few models of extra dimensions,''
[arXiv:hep-ph/0409028];
%%CITATION = HEP-PH 0409028;%%
%\cite{Landsberg:2004mj}
%\bibitem{Landsberg:2004mj}
G.~Landsberg  [CDF and D0 - Run II Collaboration]
%``Collider Searches for Extra Dimensions,''
[arXiv:hep-ex/0412028].
%%CITATION = HEP-EX 0412028;%%

%\cite{Banks:1999gd}
\bibitem{Banks:1999gd}
  T.~Banks and W.~Fischler
  %``A model for high energy scattering in quantum gravity,''
  [arXiv:hep-th/9906038].
  %%CITATION = HEP-TH 9906038;%%

\bibitem{dim} S.~Dimopoulos and G.~Landsberg,   
%''{\sl Black Holes At The LHC}'' 
Phys. Rev. Lett. {\bf 87}, 161602 (2001)
[arXiv:hep-ph/0106295];
P.C.~Argyres, S.~Dimopoulos, and J.~March-Russell,
%''{\sl Black Holes and Sub-millimeter Dimensions}'', 
Phys. Lett. B {\bf 441}, 96 (1998)
[arXiv:hep-th/9808138].

\bibitem{ehm} R.~Emparan, G.~T.~Horowitz and R.~C.~Myers,  
%">{\sl Black Holes Radiate Mainly On The Brane}"< 
Phys. Rev. Lett. {\bf 85}, 499 (2000).
 

\bibitem{Giddings3} S.~B.~Giddings and S.~Thomas, 
%">{\sl High Energy Colliders as Black Hole Factories: The End of Short Distance Physics}"< 
Phys.\ Rev.\ {\bf D 65} 056010 (2002)
[hep-ph/0106219].
 

%\cite{Hofmann:2001pz}
\bibitem{Hofmann:2001pz}
%\cite{Cheung:2001ue}
%\bibitem{Cheung:2001ue}
K.~M.~Cheung,
%``Black hole production and large extra dimensions,''
Phys.\ Rev.\ Lett.\  {\bf 88}, 221602 (2002)
[arXiv:hep-ph/0110163];
%%CITATION = HEP-PH 0110163;%%
%\cite{Uehara:2002cj}
%\bibitem{Uehara:2002cj}
Y.~Uehara
%``Virtual black holes at linear colliders,''
[arXiv:hep-ph/0205068];
%\cite{Uehara:2001yk}
%\bibitem{Uehara:2001yk}
Y.~Uehara,
%``Production and detection of black holes at neutrino array,''
Prog.\ Theor.\ Phys.\  {\bf 107}, 621 (2002)
[arXiv:hep-ph/0110382];
%%CITATION = HEP-PH 0110382;%%
%\cite{Anchordoqui:2002cp}
%\bibitem{Anchordoqui:2002cp}
L.~Anchordoqui and H.~Goldberg,
%``Black hole chromosphere at the LHC,''
Phys.\ Rev.\ D {\bf 67}, 064010 (2003)
[arXiv:hep-ph/0209337].
%%CITATION = HEP-PH 0209337;%%
%%CITATION = HEP-PH 0205068;%%
%\cite{Bleicher:2001kh}
%\bibitem{Bleicher:2001kh}

%\cite{Hossenfelder:2001dn}
\bibitem{Hossenfelder:2001dn}
S.~Hossenfelder, S.~Hofmann, M.~Bleicher and H.~St\"ocker,
%``Quasi-stable black holes at LHC,''
Phys.\ Rev.\ D {\bf 66}, 101502 (2002)
[arXiv:hep-ph/0109085].
%%CITATION = HEP-PH 0109085;%%

\bibitem{Hossenfelder:2003jz}
S.~Hossenfelder, M.~Bleicher, S.~Hofmann, J.~Ruppert, S.~Scherer and H.~St\"ocker,
%`Collider signatures in the Planck regime,''
Phys.\ Lett.\ B {\bf 575}, 85 (2003)
[arXiv:hep-th/0305262].
%%CITATION = HEP-TH 0305262;%%

%\cite{Hossenfelder:2003dy}
\bibitem{Hossenfelder:2003dy}
  S.~Hossenfelder, M.~Bleicher, S.~Hofmann, H.~St\"ocker and A.~V.~Kotwal,
  %``Black hole relics in large extra dimensions,''
  Phys.\ Lett.\ B {\bf 566}, 233 (2003)
  [arXiv:hep-ph/0302247].
  %%CITATION = HEP-PH 0302247;%%

%\cite{Casaio:2001dc}
\bibitem{Casadio:2001dc}
  R.~Casadio and B.~Harms,
  %``Black hole evaporation and compact extra dimensions,''
  Phys.\ Rev.\ D {\bf 64}, 024016 (2001)
  [arXiv:hep-th/0101154].
  %%CITATION = HEP-TH 0101154;%%

%\cite{Alexeyev:2002tg}
\bibitem{Alexeyev:2002tg}
  S.~Alexeyev, A.~Barrau, G.~Boudoul, O.~Khovanskaya and M.~Sazhin,
  %``Black hole relics in string gravity: Last stages of Hawking  evaporation,''
  Class.\ Quant.\ Grav.\  {\bf 19}, 4431 (2002)
  [arXiv:gr-qc/0201069].
  %%CITATION = GR-QC 0201069;%%
  
%\cite{Bonanno:2000ep}


\bibitem{BleicherNeu} 
%\cite{Alvarez-Muniz:2002ga}
%\bibitem{Alvarez-Muniz:2002ga}
J.~Alvarez-Muniz, J.~L.~Feng, F.~Halzen, T.~Han and D.~Hooper,
%``Detecting microscopic black holes with neutrino telescopes,''
Phys.\ Rev.\ D {\bf 65}, 124015 (2002)
[arXiv:hep-ph/0202081];
%%CITATION = HEP-PH 0202081;%%
%\cite{Mocioiu:2003gi}
%\bibitem{Mocioiu:2003gi}
I.~Mocioiu, Y.~Nara and I.~Sarcevic,
%``Hadrons as signature of black hole production at the LHC,''
Phys.\ Lett.\ B {\bf 557}, 87 (2003)
[arXiv:hep-ph/0301073];
%%CITATION = HEP-PH 0301073;% 
%%CITATION = GR-QC 0409061;%%
%\cite{Cavaglia:2004jw}
%\bibitem{Cavaglia:2004jw}
%\cite{Cavaglia:2003qk}
%\bibitem{Cavaglia:2003qk}
M.~Cavaglia, S.~Das and R.~Maartens,
%``Will we observe black holes at LHC?,''
Class.\ Quant.\ Grav.\  {\bf 20}, L205 (2003);
[arXiv:hep-ph/0305223];
%\cite{Cavaglia:2004jw}
%\bibitem{Cavaglia:2004jw}
M.~Cavaglia and S.~Das,
%``How classical are TeV-scale black holes?,''
Class.\ Quant.\ Grav.\  {\bf 21}, 4511 (2004)
[arXiv:hep-th/0404050].
%%CITATION = HEP-TH 0404050;%%%%CITATION = HEP-PH 0305223;%%

%\cite{Chamblin:2002ad}
\bibitem{Chamblin:2002ad}
  A.~Chamblin and G.~C.~Nayak,
  %``Black hole production at LHC: String balls and black holes from p p and
  %lead lead collisions,''
  Phys.\ Rev.\ D {\bf 66}, 091901 (2002)
  [arXiv:hep-ph/0206060].
  %%CITATION = HEP-PH 0206060;%%

\bibitem{Casanova:2005id}
A.~Casanova and E.~Spallucci,
%`TeV mini black hole decay at future colliders,''
Class.\ Quant.\ Grav.\  {\bf 23}, R45 (2006)
[arXiv:hep-ph/0512063].
%%CITATION = HEP-PH 0512063;%%

  
\bibitem{own2} 
S.~Hofmann, M.~Bleicher, L.~Gerland, S.~Hossenfelder, K.~Paech and H.~St\"ocker,
%``{\sl Tevatron - Probing Tev-Scale Gravity Today}'',
J.\ Phys.\ G {\bf 28}, 1657 (2002);
  S.~Hofmann, M.~Bleicher, L.~Gerland, S.~Hossenfelder, S.~Schwabe and H.~St\"ocker
[arXiv:hep-ph/0111052].
  %%CITATION = HEP-PH 0111052;%%
%\cite{Bleicher:2001kh}
 
  % M.~Bleicher, S.~Hofmann, S.~Hossenfelder and H.~Stoecker,
  %``Black hole production in 
  %large extra dimensions at the Tevatron: A  chance
  %to observe a first glimpse of TeV scale gravity,''
  %
  Phys.\ Lett.\  {\bf 548}, 73 (2002).
  %[arXiv:hep-ph/0112186].
  %%CITATION = HEP-PH 0112186;%%



%\cite{Casadio:2001wh}
\bibitem{Casadio:2001wh}
R.~Casadio and B.~Harms,
%``Can black holes and naked singularities be detected in accelerators?,''
Int.\ J.\ Mod.\ Phys.\ A {\bf 17}, 4635 (2002)
[arXiv:hep-th/0110255];
%%CITATION = HEP-TH 0110255;%%Refs to naked singularities;
I.~Ya.~Yref'eva,
% ">{\sl High Energy Scattering in the Brane-World and Black Hole Production}"<$
Part.Nucl. 31, 169-180 (2000)
[hep-th/9910269];
%\cite{Giddings:2004xy}
%\bibitem{Giddings:2004xy}
S.~B.~Giddings and V.~S.~Rychkov,
%``Black holes from colliding wavepackets,''
Phys.\ Rev.\ D {\bf 70}, 104026 (2004)
[arXiv:hep-th/0409131];
%%CITATION = HEP-TH 0409131;%%
%\cite{Rychkov:2004sn}
%\bibitem{Rychkov:2004sn}
V.~S.~Rychkov
%``Classical black hole production in quantum particle collisions,''
[arXiv:hep-th/0410041];
%%CITATION = HEP-TH 0410041;%%
%\bibitem{bf}
T.~Banks and W.~Fischler
% ">{\sl A Model For High Energy Scattering In Quantum Gravity}"<,
[arXiv:hep-th/9906038];
%\bibitem{GrWavPart}
O.~V.~Kancheli
% ">{\sl Parton picture of inelastic collisions at transplanckian energies}"<,
[arXiv:hep-ph/0208021].


%\cite{Hossenfelder:2004ze}
\bibitem{Hossenfelder:2004ze}
  S.~Hossenfelder,
  %``Suppressed black hole production from minimal length,''
  Phys.\ Lett.\ B {\bf 598}, 92 (2004)
  [arXiv:hep-th/0404232].
  %%CITATION = HEP-TH 0404232;%%


%\cite{Hossenfelder:2005bd}
\bibitem{Hossenfelder:2005bd}
  S.~Hossenfelder
  %``News about TeV-scale black holes,''
  [arXiv:hep-ph/0510236].
  %%CITATION = HEP-PH 0510236;%%

\bibitem{Hst06}
 H.~St{\"o}cker,
 Int. J. Mod. Phys. D, (2006) [arXiv:hep-ph/0605062].
 
\bibitem{Hst062}
H.~St\"ocker, to be published in Journal of Physics G (2006).
 
%\cite{Alberghi:2006qr}
\bibitem{Alberghi:2006qr}
G.~L.~Alberghi, R.~Casadio, D.~Galli, D.~Gregori, A.~Tronconi and V.~Vagnoni
%`Probing quantum gravity effects in black holes at LHC,''
[arXiv:hep-ph/0601243].
%%CITATION = HEP-PH 0601243;%%

\bibitem{cosmicrayskk} 
%\bibitem{Goyal:2000ma}
A.~Goyal, A.~Gupta and N.~Mahajan,
%``Neutrinos as source of ultra high energy cosmic rays in extra  dimensions,''
Phys.\ Rev.\ D {\bf 63}, 043003 (2001)
[arXiv:hep-ph/0005030];
%%CITATION = HEP-PH 0005030;%% 
%\cite{Emparan:2001kf}
%\bibitem{Emparan:2001kf}
R.~Emparan, M.~Masip and R.~Rattazzi,
%``Cosmic rays as probes of large extra dimensions and TeV gravity,''
Phys.\ Rev.\ D {\bf 65}, 064023 (2002)
[arXiv:hep-ph/0109287];
%%CITATION = HEP-PH 0109287;%% 
%\cite{Kazanas:2001ep}
%\bibitem{Kazanas:2001ep}
D.~Kazanas and A.~Nicolaidis,
%``Cosmic rays and large extra dimensions,''
Gen.\ Rel.\ Grav.\  {\bf 35}, 1117 (2003)
[arXiv:hep-ph/0109247].
%%CITATION = HEP-PH 0109247;%%

\bibitem{cosmicraysbh} A.~Ringwald and H.~Tu, 
%">{\sl Collider versus Cosmic Ray Sensitivity to Black Hole Production}"<, 
Phys. Lett. {\bf B 525}  135-142 (2002)
[arXiv:hep-ph/0111042]; 
J.~Feng and A.~Shapere, Phys. Rev. Lett.
88, 021303 (2002);
%\cite{Cafarella:2004hg}
%\bibitem{Cafarella:2004hg}
A.~Cafarella, C.~Coriano and T.~N.~Tomaras
%``Cosmic ray signals from mini black holes in models with extra dimensions: An
%analytical / Monte Carlo study,''
[arXiv:hep-ph/0410358];
%%CITATION = HEP-PH 0410358;%%
L.~A.~Anchordoqui, J.~L.~Feng, H.~Goldberg and A.~D.~Shapere, 
%">{\sl Black Holes from Cosmic Rays: Probes of Extra-Dimensions and New Limits on TeV-Scale Gravity}"<, 
Phys. Rev. {\bf D 65}  124027 (2002)
[arXiv:hep-ph/0112247];
%\cite{Dutta:2002ca}
%\bibitem{Dutta:2002ca}
S.~I.~Dutta, M.~H.~Reno and I.~Sarcevic,
%``On black hole detection with the OWL/Airwatch telescope,''
Phys.\ Rev.\ D {\bf 66}, 033002 (2002)
[arXiv:hep-ph/0204218];
%%CITATION = HEP-PH 0204218;%%
%\cite{Harbach:2006aj}
%\bibitem{Harbach:2006aj}
U.~Harbach and M.~Bleicher
%``No black holes at IceCube,''
[arXiv:hep-ph/0601121];
%%CITATION = HEP-PH 0601121;%%
%\bibitem{Ahn:2003qn}
E.~J.~Ahn, M.~Ave, M.~Cavaglia and A.~V.~Olinto,
%`TeV black hole fragmentation and detectability in extensive  air-showers,''
Phys.\ Rev.\ D {\bf 68}, 043004 (2003).
 

\bibitem{my} R.~C.~Myers and M.~J.~Perry,
%">{\sl Black Holes In Higher Dimensional Space-Time''}, 
Ann. Phys. {\bf 172}, 304-347 (1986).

\bibitem{hoop} K.~S.~ Thorne,
 in Klauder, J., ed., Magic without Magic, 231-258, 
(W. H. Freeman, San Francisco, 1972).
%">{\sl Black Holes In Higher Dimensional Space-Time''}, 
%Ann. Phys. {\bf 172}, 304-347 (1986).

%\cite{Voloshin:2001fe}
\bibitem{Voloshin:2001fe}
%\cite{Voloshin:2001vs}
%\bibitem{Voloshin:2001vs}
M.~B.~Voloshin,
%``Semiclassical suppression of black hole production in particle  collisions,''
Phys.\ Lett.\ B {\bf 518}, 137 (2001)
[arXiv:hep-ph/0107119];
%%CITATION = HEP-PH 0107119;%%
%``More remarks on suppression of large black hole production in particle
%collisions,''
Phys.\ Lett.\ B {\bf 524}, 376 (2002)
[arXiv:hep-ph/0111099];
%%CITATION = HEP-PH 0111099;%%
%\cite{Giddings:2001ih}
%\bibitem{Giddings:2001ih}
S.~B.~Giddings,
%``Black hole production in TeV-scale gravity, and the future of high  energy
%physics,''
in {\it Proc. of the APS/DPF/DPB Summer Study on the Future of Particle Physics (Snowmass 2001) } ed. N.~Graf,
eConf {\bf C010630}, P328 (2001)
[arXiv:hep-ph/0110127].
%%CITATION = HEP-PH 0110127;%% 

%\cite{Rychkov:2004sf}
\bibitem{Rychkov:2004sf}
V.~S.~Rychkov,
%``Black hole production in particle collisions and higher curvature gravity,''
Phys.\ Rev.\ D {\bf 70}, 044003 (2004)
[arXiv:hep-ph/0401116];
%%CITATION = HEP-PH 0401116;%%
%\cite{Kang:2004yk}
%\bibitem{Kang:2004yk}
K.~Kang and H.~Nastase
%``Planckian scattering effects and black hole production in low M(Pl)
%scenarios,''
[arXiv:hep-th/0409099].
%%CITATION = HEP-TH 0409099;%% 

\bibitem{Jevicki:2002fq}
A.~Jevicki and J.~Thaler,
%``Dynamics of black hole formation in an exactly solvable model,''
Phys.\ Rev.\ D {\bf 66}, 024041 (2002);
%[arXiv:hep-th/0203172];
%%CITATION = HEP-TH 0203172;%%
%\cite{Rizzo:2001dk}
%\bibitem{Rizzo:2001dk}
T.~G.~Rizzo,
%``Black hole production rates at the LHC: Still large,''
in {\it Proc. of the APS/DPF/DPB Summer Study on the Future of Particle Physics (Snowmass 2001) } ed. N.~Graf,
eConf {\bf C010630}, P339 (2001);
%\bibitem{Rizzo:2006uz}
T.~G.~Rizzo
%`TeV-scale black hole lifetimes in extra-dimensional Lovelock gravity,''
[arXiv:hep-ph/0601029];
%%CITATION = HEP-PH 0111230;%%
%\bibitem{Giddings2}
D.~ M.~Eardley and S.~B.~Giddings, 
%''{\sl Classical Black Hole Production in High-Energy Collisions}'', 
Phys. Rev. {\bf D 66}, 044011 (2002)
[arXiv:gr-qc/0201034];
%\cite{Rizzo:2006zb}
  T.~G.~Rizzo
  %``Noncommutative Inspired Black Holes in Extra Dimensions,''
  [arXiv:hep-ph/0606051].
  %%CITATION = HEP-PH 0606051;%%

\bibitem{Formfactors} 
%\cite{Yoshino:2002tx}
%\bibitem{Yoshino:2002tx}
H.~Yoshino and Y.~Nambu,
%``Black hole formation in the grazing collision of high-energy particles,''
Phys.\ Rev.\ D {\bf 67}, 024009 (2003)
[arXiv:gr-qc/0209003].
%%CITATION = GR-QC 0209003;%%


 
\bibitem{Solo} S.~N.~Solodukhin, 
%">{\sl Classical and quantum cross-section for black hole production in 
%particle collisions}"<, 
Phys. Lett. {\bf B 533}, 153-161 (2002) 
[hep-ph/0201248];
%\cite{Ida:2002ez}
%\bibitem{Ida:2002ez}
D.~Ida, K.~Y.~Oda and S.~C.~Park,
%``Rotating black holes at future colliders: Greybody factors for brane
%fields,''
Phys.\ Rev.\ D {\bf 67}, 064025 (2003)
[Erratum-ibid.\ D {\bf 69}, 049901 (2004)]
[arXiv:hep-th/0212108].
%%CITATION = HEP-TH 0212108;%%

\bibitem{Polchi} G.~T.~Horowitz and J.~Polchinski,
%">{\sl Instability of Spacelike and Null Orbifold Singularities}"<, 
Phys. Rev. {\bf D 66}, 103512 (2002)
[arXiv:hep-th/0206228].
 
%\cite{Landsberg:2002sa}
\bibitem{Landsberg:2002sa}
G.~Landsberg,
%``Black holes at future colliders and beyond: A review,''
[arXiv:hep-ph/0211043];
%%CITATION = HEP-PH 021104%
%\cite{Cavaglia:2002si}
%\bibitem{Cavaglia:2002si}
M.~Cavaglia,
%``Black hole and brane production in TeV gravity: A review,''
Int.\ J.\ Mod.\ Phys.\ A {\bf 18}, 1843 (2003)
[arXiv:hep-ph/0210296];
%%CITATION = HEP-PH 0210296;%%3;%%
%%CITATION = HEP-PH 0210296;%%3;%%
S.~Hossenfelder,
[arXiv:hep-ph/0412265].


\bibitem{Marcus}
    M. Bleicher, S. Hofmann, S. Hossenfelder and H. St{\"o}cker,
    Phys. Lett. {\bf{B548}}, 73 (2002).

%\cite{Chamblin:2003wg}
\bibitem{Chamblin:2003wg}
A.~Chamblin, F.~Cooper and G.~C.~Nayak,
%`Interaction of a TeV scale black hole with the quark gluon plasma at  LHC,''
Phys.\ Rev.\ D {\bf 69}, 065010 (2004)
[arXiv:hep-ph/0301239].
%%CITATION = HEP-PH 0301239;%%


\bibitem{Lonnblad:2005ah}
  L.~Lonnblad, M.~Sjodahl and T.~Akesson
%  ``{\sl QCD-supression by Black Hole Production at the LHC}'',
  [arXiv:hep-ph/0505181].
  %%CITATION = HEP-PH 0505181;%%

\bibitem{Hawk1} S.~W.~Hawking, 
%">{\sl Particle Creation by Black Holes}"< 
Comm. Math. Phys. 43, 199-220 (1975);
% ">{\sl Breakdown of Predictability in Gravitational Collapse}"<, 
Phys. Rev. D 14, 2460-2473 (1976).

%\cite{Hossenfelder:2005ku}
\bibitem{Hossenfelder:2005ku}
  S.~Hossenfelder, B.~Koch and M.~Bleicher
  %``Trapping black hole remnants,''
  [arXiv:hep-ph/0507140].
  %%CITATION = HEP-PH 0507140;%%

\bibitem{Kanti:2004nr}
P.~Kanti,
%``Black holes in theories with large extra dimensions: A review,''
Int.\ J.\ Mod.\ Phys.\ A {\bf 19}, 4899 (2004). 
 

\bibitem{Page} D.~N.~Page, 
%">{\sl Particle emission rates from a black hole: Massless particles from an uncharged, 
%nonrotating black hole}"<, 
Phys.\ Rev.\ D {\bf 13}, 198 (1976);
   R.~Casadio and B.~Harms, 
%''{\sl Black Hole Evaporation and Large Extra Dimensions}''
Phys.\ Lett. {\bf B 487}, 209-214 (2000)  
[arXiv:hep-th/0004004].



\bibitem{backreaction} %\cite{Kraus:1994by}
%\bibitem{Kraus:1994by}
  P.~Kraus and F.~Wilczek,
  %``Selfinteraction correction to black hole radiance,''
  Nucl.\ Phys.\ B {\bf 433}, 403 (1995)
  [arXiv:gr-qc/9408003];
  %%CITATION = GR-QC 9408003;%%
%\cite{Kraus:1994fj}
%\bibitem{Kraus:1994fj}
  P.~Kraus and F.~Wilczek,
  %``Effect of selfinteraction on charged black hole radiance,''
  Nucl.\ Phys.\ B {\bf 437}, 231 (1995)
  [arXiv:hep-th/9411219];
  %%CITATION = HEP-TH 9411219;%%  
 %\cite{Keski-Vakkuri:1996xp}
%\bibitem{Keski-Vakkuri:1996xp}
  E.~Keski-Vakkuri and P.~Kraus,
  %``Microcanonical D-branes and back reaction,''
  Nucl.\ Phys.\ B {\bf 491}, 249 (1997)
  [arXiv:hep-th/9610045];
  %%CITATION = HEP-TH 9610045;%%   
 %\cite{Massar:1999wg}
%\bibitem{Massar:1999wg}
  S.~Massar and R.~Parentani,
  %``How the change in horizon area drives black hole evaporation,''
  Nucl.\ Phys.\ B {\bf 575}, 333 (2000)
  [arXiv:gr-qc/9903027];
  %%CITATION = GR-QC 9903027;%%
   %\cite{Jacobson:2003wv}
%\bibitem{Jacobson:2003wv}
  T.~Jacobson and R.~Parentani,
  %``Horizon Entropy,''
  Found.\ Phys.\  {\bf 33}, 323 (2003)
  [arXiv:gr-qc/0302099];
  %%CITATION = GR-QC 0302099;%%
%\bibitem{Parikh:1999mf}
  M.~K.~Parikh and F.~Wilczek,
  %``Hawking radiation as tunneling,''
  Phys.\ Rev.\ Lett.\  {\bf 85}, 5042 (2000)
  [arXiv:hep-th/9907001].
  %%CITATION = HEP-TH 9907001;%%

%\cite{Bonanno:2006eu}
\bibitem{Bonanno:2006eu}
  A.~Bonanno and M.~Reuter,
  %``Renormalization group improved black hole spacetimes,''
  Phys.\ Rev.\ D {\bf 62}, 043008 (2000)
  [arXiv:hep-th/0002196];
  %\bibitem{Adler:2001vs}
  R.~J.~Adler, P.~Chen and D.~I.~Santiago,
  %``The generalized uncertainty principle and black hole remnants,''
  Gen.\ Rel.\ Grav.\  {\bf 33}, 2101 (2001)
  [arXiv:gr-qc/0106080];
  %\bibitem{Rizzo:2005jz}
  T.~G.~Rizzo,
  %``TeV-scale black holes in warped higher-curvature gravity,''
  [arXiv:hep-ph/0510420];
  %\cite{Rizzo:2006uz}
  %\bibitem{Rizzo:2006uz}
  T.~G.~Rizzo,
  %``TeV-scale black hole lifetimes in extra-dimensional Lovelock gravity,''
  [arXiv:hep-ph/0601029];
  %%CITATION = HEP-PH 0601029;%%
  A.~Bonanno and M.~Reuter
  %``Spacetime structure of an evaporating black hole in quantum gravity,''
  [arXiv:hep-th/0602159];
  %\bibitem{Nozari:2006bi}
  K.~Nozari and B.~Fazlpour,
  %``Thermodynamics of an evaporating Schwarzschild black hole in noncommutative
  %space,''
  [arXiv:hep-th/0605109].
  %%CITATION = HEP-TH 0605109;%%
  
  
  


 \bibitem{Daghigh:2001gy}
R.~G.~Daghigh and J.~I.~Kapusta,
%`High temperature matter and gamma ray spectra from microscopic black
%holes,''
Phys.\ Rev.\ D {\bf 65}, 064028 (2002)
[arXiv:gr-qc/0109090].


\bibitem{Humanic}
R. Barbera, B. Batyunya, Yu. Belikov, M. Botje, P. G. Cerello, A. Feliciello, T. Humanic, G. Lo Curto, 
A. Palmeri, F. Riggi,
%``Guidelines for the ALICE-ITS computer simulation.'' 
ALICE Internal Note ALICE/ITS 98-06;
T. J. Humanic, ALICE note: ALICE-INT-2005-017;
T. Humanic, B. Koch and H. St{\"o}cker, accepted for publication in Int. J. Mod. Phys. E (2006).
  
\bibitem{PYTHIA}  %\cite{Sjostrand:2001yu}
%\bibitem{Sjostrand:2001yu}
T.~Sjostrand, L.~Lonnblad and S.~Mrenna
%``PYTHIA 6.2: Physics and manual,''
[arXiv:hep-ph/0108264].
%%CITATION = HEP-PH 0108264;%  

\bibitem{CHARYBDIS} 
%\cite{Harris:2003db}
%\bibitem{Harris:2003db}
C.~M.~Harris, P.~Richardson and B.~R.~Webber,
%``CHARYBDIS: A black hole event generator,''
JHEP {\bf 0308}, 033 (2003);
[arXiv:hep-ph/0307305].
%%CITATION = HEP-PH 0307305;%%hep-ph/0307305
    
\bibitem{Atlas}
%\cite{Tanaka:2004xb}
%\bibitem{Tanaka:2004xb}
J.~Tanaka, T.~Yamamura, S.~Asai and J.~Kanzaki
%``Study of black holes with the ATLAS detector at the LHC,''
[arXiv:hep-ph/0411095].
%%CITATION = HEP-PH 0411095;%% hep-ph/0411095

\bibitem{BHex} %\cite{Harris:2004xt}
%\bibitem{Harris:2004xt}
C.~M.~Harris, M.~J.~Palmer, M.~A.~Parker, P.~Richardson, A.~Sabetfakhri and B.~R.~Webber
%``Exploring higher dimensional black holes at the Large Hadron Collider,''
[arXiv:hep-ph/0411022].
%%CITATION = HEP-PH 0411022;%%hep-ph/0411022

%\cite{Koch:2005ks}
\bibitem{Koch:2005ks}
  B.~Koch, M.~Bleicher and S.~Hossenfelder,
  %``Black hole remnants at the LHC,''
  JHEP {\bf 0510}, 053 (2005)
  [arXiv:hep-ph/0507138].
  %%CITATION = HEP-PH 0507138;%%

%\cite{Cortese:2002kf}
\bibitem{Cortese:2002kf}
  P.~Cortese {\it et al.}  [ALICE Collaboration],
  %``ALICE: Addendum to the technical design report of the time of flight system
  %(TOF),''
CERN-LHCC-2002-016.
%\href{http://www.slac.stanford.edu/spires/find/hep/www?r=cern-lhcc-2002-016}{SPIRES entry}


%\cite{Heckler:1995qq}
\bibitem{Heckler:1995qq}
  A.~F.~Heckler,
  %``On the formation of a Hawking-radiation photosphere around microscopic
  %black holes,''
  Phys.\ Rev.\ D {\bf 55}, 480 (1997)
  [arXiv:astro-ph/9601029];
  %``Calculation of the emergent spectrum and observation of primordial  black
  %holes,''
  Phys.\ Rev.\ Lett.\  {\bf 78}, 3430 (1997)
  [arXiv:astro-ph/9702027].


%\cite{Kapusta:2000xt}
\bibitem{Kapusta:2000xt}
  J.~I.~Kapusta,
  %``Relativistic Viscous Fluid Description of Microscopic Black Hole Wind,''
  Phys.\ Rev.\ Lett. 86, 1670-1673 (2001). 
  %%CITATION = ASTRO-PH 0008222;%%

  
%\cite{Csernai:1980sc}
\bibitem{Csernai:1980sc}
  L.~P.~Csernai, B.~Lukacs and J.~Zimanyi,
  %``On The Relativistic Hydrodynamical Description Of Energetic Heavy Ion
  %Reactions,''
  Lett.\ Nuovo Cim.\  {\bf 27} 111 (1980).
  %%CITATION = NCLTA,27,111;%%

%\cite{Siemens:1978pb}
\bibitem{Siemens:1978pb}
  P.~J.~Siemens and J.~O.~Rasmussen,
  %``Evidence For A Blast Wave From Compress Nuclear Matter,''
  Phys.\ Rev.\ Lett.\  {\bf 42}, 880 (1979).
  %%CITATION = PRLTA,42,880;%%

%\cite{Stoecker:1981za}
\bibitem{Stoecker:1981za}
  H.~St\"ocker, A.~A.~Ogloblin and W.~Greiner,
  %``Significance Of Temperature Measurements In Relativistic Nuclear
  %Collisions,''
%\href{http://www.slac.stanford.edu/spires/find/hep/www?irn=938653}{SPIRES entry} 
  Z.\ Phys.\ A {\bf 303}, 259 (1981).

\bibitem{shockfronts}
%\cite{Hofmann:1975by}
%\bibitem{Hofmann:1975by}
  J.~Hofmann, H.~St{\"o}cker, W.~Scheid and G.~W.,
  %``On The Possibility Of Nuclear Shock Waves In Relativistic Heavy Ion
  %Collisions,''
%\href{http://www.slac.stanford.edu/spires/find/hep/www?irn=3905756}{SPIRES entry}
{\it Report of the Workshop on BeV/nucleon Collisions of Heavy Ions: How
and Why, Bear Mountain, New York, 29 Nov - 1 Dec 1974};
%\cite{Baumgardt:1975qv}
%\bibitem{Baumgardt:1975qv}
  H.~G.~Baumgardt {\it et al.},
  %``Shock Waves And Mach Cones In Fast Nucleus-Nucleus Collisions,''
  Z.\ Phys.\ A {\bf 273}, 359 (1975);
  %%CITATION = ZEPYA,A273,359;%%
%\cite{Hofmann:1976dy}
%\bibitem{Hofmann:1976dy}
  J.~Hofmann, H.~St{\"o}cker, U.~W.~Heinz, W.~Scheid and W.~Greiner,
  %``Possibility Of Detecting Density Isomers In High Density Nuclear Mach Shock
  %Waves,''
  Phys.\ Rev.\ Lett.\  {\bf 36}, 88 (1976);
  %%CITATION = PRLTA,36,88;%%
%\cite{Stoecker:1980ppnp}
%\bibitem{Stoecker:1980ppnp}
  H.~St{\"o}cker, J.~Hofmann, J.~A.~Maruhn  and W.~Greiner,
  Prog.\ Part.\ Nucl.\ Phys.\ {\bf 4}, 133 (1980);
%\cite{Stoecker:1980vf}
%\bibitem{Stoecker:1980vf}
  H.~St{\"o}cker, J.~A.~Maruhn and W.~Greiner,
  %``Collective Sideward Flow Of Nuclear Matter In Violent High-Energy Heavy Ion
  %Collisions,''
  Phys.\ Rev.\ Lett.\  {\bf 44}, 725 (1980);
  %%CITATION = PRLTA,44,725;%%
%\cite{Stoecker:1980uk}
%\bibitem{Stoecker:1980uk}
  H.~St{\"o}cker, G.~Graebner, J.~A.~Maruhn and W.~Greiner,
  %``Hot, Dense Hadronic And Quark Matter In Relativistic Nuclear Collisions,''
  Phys.\ Lett.\ B {\bf 95}, 192 (1980);
  %%CITATION = PHLTA,B95,192;%%
%\cite{Stoecker:1986ci}
%\bibitem{Stoecker:1986ci}
  H.~St{\"o}cker and W.~Greiner,
  %``High-Energy Heavy Ion Collisions: Probing The Equation Of State Of Highly
  %Excited Hadronic Matter,''
  Phys.\ Rept.\  {\bf 137}, 277 (1986);
  %%CITATION = PRPLC,137,277;%%
%  \bibitem{Cha86}
  G.F. Chapline and A. Granik,
  Nucl.\ Phys.\ A {\bf 459}, 681 (1986);
%\bibitem{Ris90}
  D.H. Rischke, H.~St\"ocker and W.~Greiner,
  Phys.\ Rev.\ D {\bf 42}, 2283 (1990);
%\bibitem{Cas04}
 J. Casalderrey--Solana, E.V. Shuryak and D. Teaney,
 [arXiv:hep-ph/0411315];
 E.V. Shuryak, talk at at Int. Conf. Quark Matter 2005,
 Budapest, Hungary, 2005.


%\cite{Stoecker:2004qu}
\bibitem{Stoecker:2004qu}
  H.~St{\"o}cker,
  %``Collective Flow signals the Quark Gluon Plasma,''
  Nucl.\ Phys.\ A {\bf 750}, 121 (2005)
  [arXiv:nucl-th/0406018].
  %%CITATION = NUCL-TH 0406018;%%


%\cite{Satarov:2005mv}
\bibitem{Satarov:2005mv}
  L.~M.~Satarov, H.~St\"ocker and I.~N.~Mishustin,
  %``Mach shocks induced by partonic jets in expanding quark-gluon plasma,''
  Phys.\ Lett.\ B {\bf 627}, 64 (2005)
  [arXiv:hep-ph/0505245].
  %%CITATION = HEP-PH 0505245;%%
  
%\cite{Satarov:2005mv}
\bibitem{Betzinprog}
  B.~Betz et al., to be published.

%\bibitem{HorstPatent}
%  H.~St\"ocker, Deutsches Patent- und Markenamt M\"unchen,
%10 2006 007 824.1-54




\end{thebibliography}
\end{document}